\documentclass[prl,aps,twocolumn,superscriptaddress]{revtex4}
\usepackage[dvips]{graphicx}
\usepackage{txfonts}

\begin{document}
\title{Universality and backreaction in a general-relativistic  accretion of steady fluids}
\author{Janusz Karkowski}
\affiliation{M. Smoluchowski Institute of Physics, Jagiellonian University, Reymonta 4, 30-059 Krak\'{o}w, Poland}
\author{Bogusz Kinasiewicz}
\affiliation{M. Smoluchowski Institute of Physics, Jagiellonian University, Reymonta 4, 30-059 Krak\'{o}w, Poland}
\author{Patryk Mach}
\affiliation{M. Smoluchowski Institute of Physics, Jagiellonian University, Reymonta 4, 30-059 Krak\'{o}w, Poland}
\author{Edward Malec}
\affiliation{M. Smoluchowski Institute of Physics, Jagiellonian University, Reymonta 4, 30-059 Krak\'{o}w, Poland}
\affiliation{AEI MPI, Golm, Germany}
\affiliation{TPI FSU, Jena, Germany}
\author{Zdobys\l aw \'Swierczy\'nski}
\affiliation{Pedagogical University, Podchor\c a\.zych 1, Krak\'{o}w, Poland}

\begin{abstract}
The spherically symmetric steady accretion  of polytropic perfect fluids onto a black hole is the simplest flow model 
that can demonstrate the effects of backreaction. The analytic and numerical investigation reveals that backreaction
 keeps intact most of the characteristics of the sonic point. For any such system, with the  free parameter
 being the relative abundance of the fluid, the mass accretion rate achieves maximal value when the mass 
 of the fluid is  1/3 of the total mass. Fixing the total mass of the system, one observes the  existence
 of two weakly accreting regimes, one over-abundant  and the other poor  in fluid content.
\end{abstract}

\maketitle

The spherical steady accretion of gas onto a gravitational center has been investigated in the Newtonian context by
Bondi \cite{bondi} and in the Schwarzschild space-time by  Michel \cite{michel},
Shapiro and Teukolsky \cite{shapiro_teukolsky}, and others \cite{das}. The Newtonian critical accretion flow  was shown to be
stable by Balazs and others   \cite{Balazs}, while the  stability of the steady flow in the fixed Schwarzschild background
has been analyzed by Moncrief \cite{Moncrief}.
The general-relativistic description including backreaction has been formulated in \cite{malec}.
There are two reasons for inspecting steady flows in the full generality, when infalling gas
modifies the space-time geometry, that is including backreaction \cite{Wald}.
First, one can see the effects of backreaction in a simple
but nontrivial accretion model. Second, gravitational collapse of the fluid which starts as a flow dominated by
a steady accretion, can be better controlled, and that would allow one to have an insight into formation of
gravitational singularities. This paper focuses on the issue of backreaction. We show
that the backreaction strongly influences the mass accretion rate.

We will consider a spherically symmetric compact ball of a fluid falling onto a non-rotating black hole.
The black hole provides the simplest choice for the central object since one can neglect the occurrence of shock waves.
The fluid is regarded to be steady in a sense defined later
and by a black hole we understand the existence of an apparent horizon.
For the detailed derivation of equations the reader should consult \cite{malec}. Here we give only a brief description,
focusing attention on the physical assumptions. We will use  comoving coordinates
\begin{equation}
ds^2 = -N^2 dt^2 + \alpha dr^2 + R^2 \left( d \theta^2 + \sin^2 \theta d\phi^2 \right),
\label{ds}
\end{equation}
where the lapse $N$, $\alpha$ and the areal radius $R$ are functions of a coordinate radius $r$ and
an asymptotic time variable $t$. The nonzero components of the extrinsic curvature $K_{ij}$ of
the $t = \mathrm{const}$ slices read $K_r^r = \frac{1}{2N\alpha} \partial_t \alpha, \;\;
K_\phi^\phi = K_\theta^\theta = \frac{\partial_t R}{NR} = \frac{1}{2} ( \mathrm{tr} K - K_r^r)$.
Here $\mathrm{tr} K = \frac{1}{N} \partial_t \ln \left( \sqrt{\alpha} R^2 \right)$.
The mean curvature of two-spheres of constant radius $r$, embedded in a Cauchy hypersurface is
 $p = \frac{2 \partial_r R}{R \sqrt{\alpha}}$. The energy-momentum tensor of the  perfect fluid reads
 $T^{\mu\nu} = (\tilde p + \varrho) u^\mu u^\nu + \tilde p g^{\mu\nu}$,
 ($u_\mu u^\mu = -1$) where $u^\mu$ denotes the four-velocity of the fluid,
 $\tilde p$ is the pressure and $\varrho$ the energy density in the comoving frame.

The areal velocity $U$ of a comoving particle designated by coordinates $(r,t)$
 is given by $U(r,t) = \frac{\partial_t R}{N} = \frac{R}{2}(\mathrm{tr} K - K_r^r)$.
 From the Einstein constraint equations one has
\begin{equation}
pR = 2 \sqrt{1 - \frac{2m(R)}{R} + U^2}
\label{p}
\end{equation}
and $\partial_R (R^2 U) - R^2 \mathrm{tr} K = 0$. We use here and in what follows
the relation $\partial_r = \sqrt{\alpha}(pR/2)\partial_R$ in order to eliminate
the differentiation with respect the comoving radius $r$. The quasilocal mass
$m(R)$ obeys the equation $\partial_R m(R) = 4 \pi R^2 \varrho$. The mass
accretion rate is $\dot m = (\partial_t - (\partial_t R)\partial_R) m(R) =
- 4 \pi NR^2U (\varrho + \tilde p) $. A more familiar form of that is
$\dot m = -4 \pi R^2 n U$, where $n$ is the baryonic density (see below).
A standard condition for the steady collapsing fluid is that all its
characteristics are constant at a fixed $R$ (see \cite{Courant}). In analytical terms
$\partial_t X|_{R = \mathrm{const}} = (\partial_t - (\partial_t R) \partial_R) X = 0$,
 where $X = \varrho, U, a \dots$ \cite{Misner}. The Einstein evolution equation
 $\partial_t U = p^2 R^2 \partial_RN/4 - m(R)N/R^2 - 4 \pi N R \tilde p$ and
 the energy conservation equation $\partial_t \varrho = -N \mathrm{tr} K (\varrho + \tilde p)$
 become ordinary differential equations with respect $R$, assuming steady flow.
  Selfgravitating steady fluids still allow for the change in time of some geometric
   quantities, like the mean curvature $p$ or the area of the black hole.
   It is easy to show that for a steady flow $\partial_R \dot m = 0$ \cite{malec}.

By a black hole is meant a region within an apparent horizon to the future, i.e.,
a region enclosed by an outermost sphere $S_A$ on which the optical scalar $\theta_+
\equiv \frac{pR}{2} + U$ vanishes \cite{nom}. (The other condition, that
 $\theta_-\equiv \frac{pR}{2} - U > 0$ for all spheres outside $S_A$, is
  satisfied trivially for a steady accretion.) That means that on $S_A$
  the ratio $2m_{BH}/R_{AH}$ becomes 1, where $R_{AH}$ is the areal radius
   of the apparent horizon and $m_{BH}\equiv m(R_{AH})$ is the mass of the black hole.
    We shall specialize to polytropic perfect fluids
   $\tilde p = K \varrho^\Gamma$, with $\Gamma$ being a constant, ($1 \le \Gamma \le 5/3$).

Furthermore, assume the radius $R_\infty$ of the ball of fluid and boundary data are
such that $|U_{\infty}| \ll \frac{m(R_\infty)}{R_\infty} \ll a_\infty$.
 These boundary conditions are needed in order to glue the steady fluid
 with the external Schwarzschild geometry (see a discussion later).
 The momentum conservation equation $\nabla_\mu T^\mu_r = 0$ ---
in comoving coordinates
 \begin{equation}
\label{euler}
N \partial_R \tilde p + (\tilde p + \varrho) \partial_R N = 0
\end{equation}
--- can be integrated, yielding (with $N(R_\infty) = 1$)
\begin{equation}
a^2 = - \Gamma + \frac{\Gamma +a^2_\infty}{N^\kappa}.
\label{bernoulli}
\end{equation}
Here $\kappa = \frac{\Gamma - 1}{\Gamma}$. (\ref{bernoulli}) can be regarded as
 the general-relativistic version of the Bernoulli equation. Another useful relation that follows from (\ref{euler}) is
$N=\frac{Cn}{\tilde p + \varrho}$.
Here appears the baryonic number density
 $ n = C \exp \int_{\varrho_\infty}^\varrho ds \frac{1}{s + \tilde p(s)}. $
The baryonic number density works, for perfect barotropes, as an integration factor;
 if $T_{\mu\nu}$ is conserved, then $\nabla_\mu \left( n u^\mu \right) = 0$.
 This representation   of the lapse function shows that the lapse $N$
 is constant along an orbit of the constant areal radius $R$.

In the general-relativistic case the sonic point is defined as a location where
the length of the spatial velocity vector $|{\vec U}| = |U|/(pR/2)$ equals $a$.
 Therefore at a sonic point $|U| = \frac{1}{2}pRa$. In the Newtonian limit this
  coincides with the standard requirement $|U|=a$. In the following we will denote
   by the asterisk all values referring to the sonic points, e.g. $a_\ast$, $U_\ast $, etc.
   The four characteristics, $a_\ast$, $U_\ast$, $m_\ast / R_\ast$ and $c_\ast$ are related
   \cite{malec}, $a_\ast^2 \left( 1 - \frac{3m_\ast}{2R_\ast} + c_\ast \right) = U_\ast^2 =
   \frac{m_\ast}{2R_\ast} + c_\ast$, where $c_\ast = 2 \pi R^2_\ast \tilde p_\ast$. The infall velocity $U$ reads
\begin{equation}
U=U_\ast \frac{R^2_\ast}{R^2}
\left(  \frac{1 + \frac{\Gamma}{a^2}}{1 + \frac{\Gamma}{a^2_\ast} } \right)^{1/(\Gamma -1)}.
\label{U}
\end{equation}
Here $U_\ast$ is the negative square root. From the relation between the pressure and the
energy density, one obtains, using  equation (\ref{bernoulli})
\begin{equation}
\varrho = \varrho_{\infty } \left( a/a_\infty \right)^{2/(\Gamma - 1)} = \varrho_\infty
\left( - \frac{\Gamma}{a_\infty^2} + \frac{\frac{\Gamma}{a_\infty^2}  + 1}{N^\kappa} \right) ^ \frac{1}{\Gamma - 1},
\label{rho}
\end{equation}
where the constant  $\varrho_\infty$ is equal to the mass density of a collapsing
fluid at the boundary $R_\infty$. The steady fluid is described by equations
(\ref{bernoulli} -- \ref{rho}). They constitute an integro-algebraic system of equations,
with a bifurcation point at the sonic point, where two branches (identified as accretion
or wind) do cross; that is a well known feature of that problem, present also in models
 with a test fluid \cite{bondi} -- \cite{Balazs}, \cite{malec}. That requires some caution in doing
 numerics and a careful selection of the solution branch. Notice that these equations
  are expressed exclusively in terms of quantities that are steady in the sense of the former definition.

This  collapsing compact cloud can be connected  to the Schwarzschild  exterior.
The ideal model of a steady flow would be that engineered by a "daemon" who just fixes boundary data conditions;
a similar picture  might well be valid in some binaries.
Alternatively, one can imagine an initial configuration with a steady fluid annulus
 around the black hole and an external transition zone that is not steady (since both
 pressure and  the mass density must be made to vanish there) and that can expand outward
 and inward with the speed of sound, perturbing the external layers of the steady interior.
 The boundary can be connected at $R_\infty$, via an almost massless transition zone, to the
 external Schwarzschild geometry; the standard junction conditions --- continuity of
  the metric and of the transversal extrinsic curvature, and the condition $N_\infty =
  1$ --- require now the appropriate choice of the slicing of the exterior region.
 We verified in a number of numerical calculations that for short evolution times
 ($t \ll R_\infty / a_\infty$) and under the former boundary
 conditions the bulk of the fluid is steady. (Notice, that this observation indicates that
such initial data can be useful in order to model the gravitational collapse.)
 The asymptotic mass $m$ of the spacetime would
   be well approximated by $m_\infty$, the sum of the masses of the black hole
    and the quasi-stationary gas, for a suitably chosen transition zone.

    In numerical calculations it will be convenient to represent the mass $m(R)$ in the form
\begin{equation}
m(R) = m - 4 \pi \int_R^{R_\infty} dr r^2 \varrho.
\label{mR}
\end{equation}
Notice that $m(\infty) = m_\infty \approx m$. Assuming steady flow, one can
obtain following expression for the lapse $N$ \cite{malec}
\begin{eqnarray}
N & = & \frac{pR}{p(R_\infty) R_\infty} \beta(R),
\nonumber\\
\beta (R) & = &  \exp \left( - 16 \pi \int_R^{R_\infty} \left(  \tilde p
 + \varrho \right) \frac{ds}{p^2 s} \right).
\label{Nb}
\end{eqnarray}
The full description of the steady accretion that will be discussed later is
 given by equations (\ref{p}), (\ref{mR}), (\ref{bernoulli}), (\ref{U}) and 
 (\ref{Nb}); this is equivalent to the former system.

One of   two main results of this paper is realization that significant information about
the full system with backreaction can be obtained through the investigation
of steady flows with the backreaction being ignored.

Namely, \textbf{let be given $   \Gamma \in (1, 5/3]$, asymptotic data
 $\varrho_\infty$ and $a_\infty$ and let $|U_\infty| \ll \frac{m(R_\infty)}{R_\infty} \ll a_\infty$. 
 Assume that the sonic point is located outside the apparent horizon, i.e., $U^2_\ast < 0.25$.
  Then the  sonic point parameters $a^2_\ast$, $U^2_\ast$ and $m_\ast / R_\ast$ 
  in the above model with backreaction are the same as in the test fluid  accretion 
  with the same asymptotic data \cite{remark}.}

\textbf{Sketch of the reasoning}.
We will apply the bootstrap-type argument. Let us assume temporarily that $U^2_\ast  < 0.1$,
 that is $R_\ast > 5 m_\ast$. This implies $pR/2 \ge \sqrt{0.7}$, $a^2 \le a^2_\ast \le 1/7$
 and $16\pi\int_R^{R_\infty} (\tilde p + \varrho) \frac{1}{p^2 s} ds \le 80 \frac{m - m(R_\ast)}{49 R_\ast}$
 for $R\ge R_\ast $.
  
If $m - m(R_\ast) \le (49/80) m(R_\ast)$, then the integrand present
 in the exponent of the function $\beta (r)$ is smaller than $m_\ast / R_\ast$.
 That in turn means  that $N \ge 
 \sqrt{1- \frac{4m_\ast}{R_\ast} + U^2_\ast}$ and
\begin{equation}
a^2_\ast \le - \Gamma + \frac{\Gamma + a^2_\infty}{{\sqrt{1 - \frac{4m_\ast}{R_\ast} + U^2_\ast}}^\kappa}.
\label{estimate}
\end{equation}
Solutions of this inequality are bounded from above by the solution of the equation 
\begin{equation}
\tilde a^2  = - \Gamma + \frac{\Gamma + a^2_\infty}{{\sqrt{1 - 
\frac{2 \tilde m}{R_\ast} + \tilde U^2}}^\kappa},
\label{estimate.}
\end{equation}
where $\tilde m = 2 m_\ast$, at the point where $\tilde a^2 / \left(1 +
3 \tilde a^2 \right) = \tilde m / (2 R_\ast) = \tilde U^2$. 
Note that $R_\ast > 2 \tilde m $; this allows us to use the estimate of Theorem 2 of \cite{malec}. Thus one has
\begin{equation}
a^2_\ast < \tilde a^2 \le \frac{2a^2_\infty}{5 - 3 \Gamma + \frac{3 \tilde a^2 
(\Gamma - 1)^2 (9\Gamma - 7)}{4 \Gamma \left(1 + 3 \tilde a^2 \right)}}.
\label{theorem2}
\end{equation}
Inserting this  estimate into (\ref{rho}), one can bound the matter density,
 $\varrho(R) \le \varrho_\infty \left( 1 + C(\Gamma, a_\ast) / \left( 5 - 3 
 \Gamma + a_\ast^2 \right) \right)$. Here the constant $C$ is of the order of 10. 
 This in turn allows one to replace the bound on the exponent of $\beta$ by 
 a much better one of the order $\left( m_\ast / R_\ast \right) \left( R_\ast /R_\infty \right)$;
 this is clearly negligible in comparison to  $m_\ast / R_\ast$ (save for a small region
   around $\Gamma = 5/3$) and one can repeat  the previous argument, but this time 
   assuming $\beta = 1$. This would lead to the improved estimate of $a$ and $\varrho$ outside $R = 5R_\ast$.
    Further refinement --- the extension into bigger values of $U^2_\ast$ and of $m_\ast/m$, 
    and to all $\Gamma$'s in the interval $(1, 5/3]$ --- would be done by splitting
    the integral of $\beta$ into two integrals, $\int_{R_\ast}^{R_\infty } \dots =
     \int_{R_\ast}^{5R_\ast} \dots + \int_{5R_\ast }^{R_\infty } \dots$ and separately estimating each term.
  
Using the preceding estimates, one gets
\begin{eqnarray}
c_\ast & = & \frac{a^2_\ast}{\Gamma} 2 \pi R^2_\ast \varrho_\ast \le C a^2_\ast 
\frac{m_\ast}{R_\infty} \frac{R_\ast^2}{R_\infty^2} \ll \frac{m_\ast}{R_\ast}.
\end{eqnarray}
That implies that both $c_\ast$ and the exponent of $\beta$ can be put to zero 
in all relations between characteristics of the sonic point. Inserting 
this information into the Bernoulli equation (\ref{bernoulli}) one obtains that at a sonic point \cite{malec}
\begin{equation}
1 + y (3 \Gamma - 1) = 3 \left(a^2_\infty + \Gamma \right) y^\frac{\Gamma + 1}{2 \Gamma},
\label{sonic}
\end{equation}
where $y = 1 - 3 m_\ast / (2R_\ast)$. Coefficients of this algebraic equation do not depend on 
the asymptotic mass density and therefore $y$ is $\varrho_\infty$-independent. 
The sonic mass $m_\ast$ and the sonic radius $R_\ast$   clearly depend  on $\varrho_\infty$ but their
 ratio is constant. In fact, $m_\ast / R_\ast$ must be exactly the same as in the case when 
 the backreaction can be neglected, that is when the mass of the fluid outside the black hole is small 
 in comparison to the total mass. The same conclusion holds true also for other parameters of the sonic 
 point, the fluid velocity $U_\ast$ and the speed of sound $a_\ast$. In conclusion: $a^2_\ast $, $U^2_\ast $
 and $m_\ast / R_\ast$ can be inferred from a suitable steady flow with a test fluid.

It follows from the above discussion that the mass density changes moderately outside the sonic point. 
In fact, for $R \in [R_\ast, 10 R_\ast]$ $\varrho(R) \le C(R) \varrho_\infty$ with $C(R_\ast)$ being of 
the order of 10 and $C(10 R_\ast)$ being slightly larger than 1. Assuming that $R_\infty \gg 10 R_\ast$ 
one can show (taking into account the fact, that the region $(10 R_\ast, R_\infty)$ has a much larger 
volume than $(R_\ast, 10R_\ast)$) that under the preceding assumptions the mass of outer layers of 
the fluid $m - m_\ast = 4 \pi \int_{R_\ast}^{R_\infty} dr r^2 \varrho = \gamma \varrho_\infty$ 
with a constant $\gamma$ determined in practice only by $R_\infty$. Alternatively, $m_\ast / m =
 1 - \varrho_\infty (\gamma / m)$; the ratio $m_\ast / m$ is a linearly decreasing function of $\varrho_\infty$.

The  rate of  the  mass accretion $\dot m$ within the steadily
accreting fluid can be expressed as below (see Eq. (6.1) in \cite{malec})
\begin{equation}
\dot m = \pi m_\ast^2 \varrho_\infty \frac{R^2_\ast}{m_\ast^2} \left( \frac{a_\ast^2}{a_\infty^2} 
\right)^\frac{(5 - 3 \Gamma)}{2(\Gamma - 1)} \left(1 + \frac{a^2_\ast}{\Gamma } \right) \frac{1 + 3 a_\ast^2}{a^3_\infty}.
\label{dotm}
\end{equation}
The whole dependence on $\varrho_\infty$ is contained in  the factor $m_\ast^2 \varrho_\infty$. 
This achieves (fixing $m$) a maximum at $m_\ast = 2 m / 3$ and tends to zero at both ends: 
i) $m_\ast \to m$ (when the density $\varrho$ tends to zero) and ii) $m_\ast / m \to 0$ 
(when the mass of black hole is negligible in comparison to the mass of the fluid).
 In this place we invoked the assumptions concerning boundary conditions.

  Thus, one arrives at our second main result: \textbf{Amongst steadily accreting systems 
  of the same  $R_\infty$, $a_\infty$ and $m$ those will be most efficient for which $m_\ast = 2 m / 3$.}
   The factor 2/3 is universal --- independent of the parameters $R_\infty$, $\Gamma$ and $a_\infty$.

Since the mass of the fluid  $m_f$ is close  to $m-m_\ast $  and $m=m_f+m_{BH}$, the above
means that the maximum of the mass accretion takes place when the mass of the fluid is
a half of the mass of the black hole.
   We would like to point that in (\ref{dotm}) the mass $m_\ast $ is  only approximately
   constant, and hence $\dot m$ is not strictly constant in time.
The fulfilment of $\dot m \tau \ll m_\ast$,
    where $\tau $ is the evolution time, can be understood as yet another criterion for the validity
    of the steady flow approximation.

In the remaining part of the paper the asymptotic mass $m$ has been   normalized to 1.
In order to test the above results, we performed a series of numerical calculations,
 according to the following scenario. i)  Fix a value of the adiabatic index $\Gamma $
 and the asymptotic speed of sound $a^2_\infty$; change in a systematic way the asymptotic
  mass density. ii) For the same $\Gamma$ choose new $a^2_\infty$ and change the asymptotic 
  mass density. iii) Choose new $\Gamma$ and repeat the preceding procedure. 
  Below we shall describe only a small subset of the numerical evidence.

\begin{table}[ht]
\caption{Characteristics of the sonic point $a_\ast^2$, $|U_\ast|$, $R_\ast$ for the $\Gamma
 = 4/3$ polytrope and $a_\infty^2 = 0.1$. The last column shows the areal radius of 
 the apparent horizon. The mass accretion rate (second column) achieves maximum at $m_{BH} \approx 0.663$.}
\begin{tabular}{cccccc}
$m_f / m$ & $\dot m$ & $R_\ast$ & $U_\ast$ & $a_\ast^2$ & $R_{AH}$ \\
\hline
\hline
$4.186\times 10^{-32}$    & $4.156\times 10^{-48}$  & $4.290$ & $0.34137$  & $0.179184$   & $1.993$\\
$4.186\times 10^{-17}$    & $4.156\times 10^{-33}$  & $4.290$ & $0.34137$  & $0.179184$   & $1.993$\\
$0.041$                   & $3.815\times 10^{-18}$  & $4.110$ & $0.341369$ & $0.179182$   & $1.906$\\
$0.083$                  & $6.978\times 10^{-18}$  & $3.931$ & $0.341371$ & $0.179183$   & $1.821$\\
$0.167$                  & $1.152\times 10^{-17}$  & $3.571$ & $0.341371$ & $0.179182$   & $1.656$\\
$0.251$                   & $1.397\times 10^{-17}$  & $3.212$ & $0.341371$ & $0.179183$   & $1.492$\\
$0.333$                 & $1.469\times 10^{-17}$  & $2.859$ & $0.341371$ & $0.179183$   & $1.326$\\
$0.418$                  & $1.403\times 10^{-17}$  & $2.493$ & $0.341399$ & $0.179184$   & $1.156$\\
$0.628$                   & $8.612\times 10^{-18}$  & $1.594$ & $0.341369$ & $0.179182$   & $0.740$\\
$0.837$                 & $2.187\times 10^{-18}$  & $0.696$ & $0.341394$ & $0.179185$   & $0.322$\\
$0.879$                  & $1.264\times 10^{-18}$  & $0.516$ & $0.341399$ & $0.179184$   & $0.239$\\
$0.962$                   & $1.277\times 10^{-19}$  & $0.156$ & $0.34137$  & $0.179187$   & $0.072$\\
$0.983$                  & $2.382\times 10^{-20}$  & $0.067$ & $0.341399$ & $0.179185$   & $0.031$\\
\end{tabular}
\end{table}

It appears that the characteristics $a_\ast^2$, $U_\ast^2$, $m_\ast/R_\ast$ of the sonic point 
{\it practically} do not depend, for a given $\Gamma$ and $a^2_\infty$, on the energy density 
$\varrho_\infty$ (exemplary results are shown in Table 1), suggesting a precision of $10^{-5}$. 
The more reliable measure of the numerical error is the difference between the mass $m = 1$ 
(assumed in the equations) and the mass found numerically. Our results (see Table 1) suggest
that numerical error is smaller than 0.5 \%. The quantity $c_\ast$ is negligible in comparison 
to other sonic point parameters. The mass $m_\ast$ behaves like $m_\ast = 1 - 
\gamma \varrho_\infty$ (Fig. 1). And finally, the extremum of $\dot m$ is achieved when 
$m_\ast \approx 2/3$ (Fig. 2). Notably, numerical results extend the regime in which the 
universality is observed into fluids characterized by the speed of sound $a_\infty$ exceeding 1. 
This is remarkable, since that implies $R_\ast < R_{AH}$ and the theoretical analysis 
presented above would not be valid.
\begin{figure}[h]
\includegraphics[width=8cm]{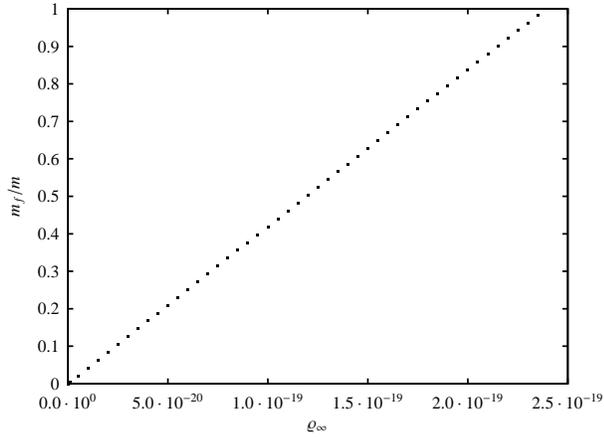}
\caption{The dependence of $\frac{m_f}{m} \approx 1 - m_\ast$ on the asymptotic mass density.}
\end{figure}
%%%%%%%%%%%%%%%%%%%%%%%%%%%%%%%%%%%%%%%%%%%%%%%%%%%%%%%%%%%%%%%%%%%%%%%%
%%%%%%%%%%%%%%%%%%%%%%%%%%%%%%%%%%%%%%%%%%%%%%%%%%%%%%%%%%%%%%%%%%%%%%%%
\begin{figure}[h]
\includegraphics[width=8cm]{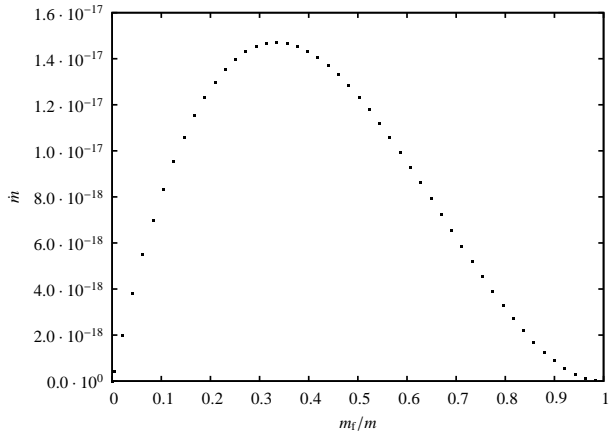}
\caption{The dependence of $\dot m$ on the ratio $\left(m - m_\ast \right) / m$.}
\end{figure}
%%%%%%%%%%%%%%%%%%%%%%%%%%%%%%%%%%%%%%%%%%%%%%%%%%%%%%%%%%%%%%%%%%%%%%%%

Formula (\ref{dotm}) implies that the mass accretion rate  in the case with backreaction
(when $m_\ast < m$) can be only smaller from the mass accretion rate without backreaction 
(when $m_\ast = m$), with the same asymptotic parameters $\rho_\infty$, $a_\infty$, $m$.
Numerical results    confirm the preceding argument that   $\gamma $ depends
essentially only  on the size of support.

In the case  of polytropes with   $\tilde p = K n^\Gamma$
 the Bernoulli equation  reads $N \left( \Gamma - 1 - a^2_\infty \right) = \Gamma - 1 -a^2$.
The analytic arguments of the type presented above can be applied and conclusions are the same.
   Similar results should be true for Newtonian
  massive fluids  for  $\Gamma $   significantly smaller than $5/3$.

In conclusion, in the simple model of accretion with backreaction considered here, one can 
get all parameters describing the sonic point, with the exception of its location $R_\ast$
and mass $m_\ast$ simply from a related polytropic model with the test fluid having the same 
index $\Gamma$ and the same asymptotic speed of sound $a_\infty$. The main result
 is that the mass accretion rate  $\dot m$ achieves maximum
at $m_f/m_{BH}\approx 1/2$;
therefore there exist two different regimes, $m_f/m_{BH}\ll 1$ and $m_f/m_{BH}\gg 1 $, with low accretion.
In contrast to that, in the test fluid approximation the quantity $\dot m$ grows with $\varrho_{\infty }$
(that is with  $m_f/m_{BH}$).
The mass accretion rate $\dot m$ sets the upper limit for the luminosity of the system. Hence  this simple
analysis of accretion suggests the  existence of two  weakly luminous    regimes:
one   rich in fluid  with $m_f/m_{BH}\gg 1$  and the other with a small amount
of fluid, $m_f/m_{BH}\ll 1$. We noticed also   that the normalized
efficiency of accretion (defined as  $\dot m / m_{BH}$) goes to zero.

Results of this paper show the importance of the backreaction.    We believe  that the qualitative
features demonstrated by the spherically symmetric model --
the dependence of the mass accretion on $  m_f / m_{BH}$ and
the existence of two weakly  accreting regimes --  will appear in the  descriptions
of    accreting systems with angular momentum.

Acknowledgements. EM thanks Bernd Schmidt for discussions.
This paper is partially supported by the MNII grant 1PO3B 01229.


\begin{thebibliography}{99}

\bibitem{bondi} H. Bondi, \textit{Mon. Not. R. Astron. Soc.} \textbf{112}, 192(1952).
\bibitem{michel} F. C. Michel, \textit{Astrophys. Space Sci.} \textbf{15}, 153(1972).
\bibitem{shapiro_teukolsky} S. Shapiro and S. Teukolsky,  Black Holes, White Dwarfs and Neutron Stars, Wiley, New York, 1983.
\bibitem{das} T.K. Das, \textit{Astron. Astrophys.} {\bf 374}, 1150(2001); B. Kinasiewicz and T. Lanczewski,
\textit{ Acta Phys. Pol.} {\bf B36}, 1951(2005).
\bibitem{Balazs} N. L. Balazs, {\it Mon. Not. R. Astr. Soc.} {\bf 160}, 79(1972); R. F.  Stellingwerf  and J. Buff, {\it Astrophys. J.}
{\bf 198}, 671(1978); A. R. Garlick, {\it Astr. Astrophysics}, {\bf 73}, 171(1979); J. Patterson, J. Silk and J. P. Ostriker,
{\it Mon. Not. R. Astr. Soc.}, {\bf 191}, 571(1980).
\bibitem{Moncrief}    V. Moncrief, {\it  Astrophys. J.} {\bf 235}, 1038(1980).
\bibitem{Wald} R. Wald, General Relativity, Chicago, University of Chicago Press, 1984.
\bibitem{malec} E. Malec, \textit{ Phys. Rev.} \textbf{D60}, 104043 (1999).
\bibitem{Courant} R. Courant and K. O. Friedrichs, Supersonic Flow and Shock Waves, Appl. Math. Sc. {\bf 21}, Springer-Verlag.
\bibitem{Misner} Notice that $(\partial_t - (\partial_t R)\partial_R) \propto \partial_T$, where $T$ is
the Misner-Sharp time, see: C. W. Misner and D. H. Sharp, \textit{Phys. Rev.} \textbf{136B}, 571(1964).
\bibitem{remark} In the test fluid approximation the parameters of the sonic point do not depend on
the central mass. See Theorem 2 and Eqs. (5.10 -- 5.11) in \cite{malec}.
\bibitem{nom} E. Malec and N. O'Murchadha, \textit{Phys. Rev.} \textbf{D50}, 6033(1994).

\end{thebibliography}
\end{document}